\documentclass[prd,aps,twocolumn,preprintnumbers,showpacs,nofootinbib,superscriptaddress,notitlepage]{revtex4-1}

\usepackage{hyperref}

\usepackage{bm}
\usepackage{amsmath}
\usepackage{slashed}
\usepackage{xcolor}
\usepackage{graphicx}
\usepackage{caption}
\usepackage{subcaption}
\usepackage[normalem]{ulem}
\usepackage[utf8]{inputenc}
\usepackage[T1]{fontenc} % if needed

\newcommand{\beq}{\begin{eqnarray}}
\newcommand{\eeq}{\end{eqnarray}}

\begin{document}

\title{Why is LaMET an effective field theory for partonic structure?}

\author{Xiangdong Ji}
\email{xji@umd.edu}
\affiliation{Center for Nuclear Femtography, SURA, 1201 New York Ave. NW, Washington, DC 20005, USA}
\affiliation{Maryland Center for Fundamental Physics,
Department of Physics, University of Maryland,
College Park, Maryland 20742, USA}
%\affiliation{Tsung-Dao Lee Institute, Shanghai Jiao Tong University, Shanghai, 200240, China}

\date{\today{}}

\begin{abstract}
Partons are effective degrees of freedom describing the structure of hadrons
involved in high-energy collisions. Familiar theories of partons are
QCD light-front quantization and soft-collinear effective theory,
both of which are intrinsically Minkowskian and appear unsuitable for
classical Monte Carlo simulations. A ``new'' form of the parton theory
has been formulated in term of the old-fashioned, Feynman's infinite momentum
frame, in which the parton degrees of freedom are filtered
through infinite-momentum external states. The partonic
structure of hadrons is then related to the matrix elements of
static (equal-time) correlators in the state $|P^z=\infty\rangle$. This
representation lays the foundation of large-momentum effective theory (LaMET)
which approximates parton physics through a systematic $M/P^z$ expansion
of the lattice QCD matrix elements at a finite but large momentum $P^z$,
and removes the residual logarithmic-$P^z$ dependence by the
standard effective-field-theory matching and running.

\end{abstract}
\maketitle

\section{Introduction}

In 2013, I wrote a paper on ``Parton Physics on Euclidean Lattice''\cite{Ji:2013dva}, describing
a new method to directly calculate parton distribution functions (PDFs)
and other parton observables using Euclidean quantum chromodynamics (QCD),
which can be implemented in lattice field theory. This paper has generated much interest in the
community of parton physics: many follow-up works have been published, and
a few reviews have appeared describing the rapid progress in both theory and lattice QCD
simulations~\cite{Cichy:2018mum,Ji:2020ect}.

In 2014, I realized that the method in fact implies an effective
field theory (EFT) approach to calculating partonic structure of a hadron,
allowing a controlled systematic approximation to almost any parton properties.
Therefore, I wrote
another paper explaining the basic principle behind the previous one~\cite{Ji:2014gla}.
This paper did not get much attention, at least judged by the number of citations.
A more fundamental reason perhaps is that many, including my friends,
may consider this to be a dubious EFT, therefore they have
kept silent about this, quoting my method only as the ``quasi-PDF'' approach.
A few critics either don't care, or only raise questions through referee reports.
This is understandable because EFT is nowadays synonymous to ``systematic and
fundamental'', differing from uncontrolled ``models''.
I understand perfectly well that the word should not be abused.
Otherwise, we will have more EFTs than the number of theorists in the field.
Actually, a long listing of EFTs is given in a website
by I. Stewart for his famous online course on EFT~\cite{Stewart:2020aa}
and also~\cite{Manohar:2018aog,Davidson:2020xfi}.
A former smart postdoc of mine has done a lot of important work
in this new field, but never mentioned LaMET in his papers. Troubled
by this, I finally ask him why not quoting my LaMET paper? He honestly
replied, ``I never understand why it is an EFT. Where is the
effective Lagrangian?''

This is indeed a good question! LaMET reverses the logic of a
standard EFT and use the full QCD Lagrangian as the effective one.
It follows from a ``new'' insight that partons can be generated by an
infinite-momentum external state. It seeks an approximate solution through a
systematic expansion using finite but large momentum, similar to that
discrete points have been used to approximate the continuous space-time in lattice QCD.
Why I still think of it as an effective theory has been explained in a recent review paper
I co-authored~\cite{Ji:2020ect}.  However, the reivew is long and many people don't
have the patience to read it through, except some poor graduate students
whose advisors assign it as a reading material. On the other hand,
perhaps only a student with good training in quantum field theory (QFT) can
finally work through the logic in the paper.

In my view, LaMET provides, for the first time, a practical and
systematic theoretical framework to calculate parton physics
through Euclidean lattice QCD simulations, a goal
that late Ken Wilson and others tried to achieve through directly solving
Minkowskian light-front QCD~\cite{Wilson:1994fk,Brodsky:1997de}, a well-known theory
of partons. In fact, recent works on transverse-momentum-dependen(TMD) parton distributions
demonstrate this point quite clearly~\cite{Ji:2018hvs,Ebert:2019okf,Ji:2019ewn,Vladimirov:2020ofp}.
The upcoming Electron-Ion Collider~\cite{Accardi:2012qut} will allow to measure
a very broad range of observables with unprecedented accuracy. LaMET may provide the
unique tool to link most of these unambiguously to the partonic
structure of the proton within QCD.

For all reasons above, I wrote up this article based on a seminar I gave.
The article aims at beginning graduate students, perhaps after two semesters
of a QFT course. All comments are welcome for a modified version,
including missing references.

\section{What is an Effective Theory?}

When taking my first physics course in middle school, I remembered
one comment from a teacher very clearly: Physicists always consider idealized
concepts so that a problem can be simplified to a point that it has a simple solution.
Hence we have ``point'' particles in Newton's mechanics, ``ideal'' gases in thermodynamics,
and ``ideal'' fluid in fluid dynamics, etc. These simplifications
make the main physics points clear and fun.

As an undergraduate, after learning calculus, I suddenly realized that
this is called Taylor expansion. Imagine some physics quantity
$f(x,\epsilon,\delta,...)$ depending on the variable $x$ and many
other parameters $\epsilon$, $\delta$,... etc, (e.g.,
the radius of the Earth when studying its rotation around the Sun),
one can simplify the problem by expanding around the ideal limit, $\epsilon=\delta...=0$,
\begin{eqnarray}
    f(x,\epsilon,\delta,...) &&= f(x, 0,0,...)+ \epsilon f_\epsilon(x,0,0,...)\nonumber\\
        && +\delta f_\delta (x,0,0,...)+... \ .
\end{eqnarray}
The first term is what we learn in middle and high schools, and many
frontier researches are about understanding high-order
terms in the series. This way of doing physics may be called an effective
approach. So an effective theory is in a certain sense about
a Taylor expansion, which may lead to Nobel prizes
in precision measurements, e.g., the magnetic moment of the electron,
the pulse-period variation in a neutron star binary, etc.

There are many such examples in college physics. The most famous one
is probably the multiple expansion in electrostatics. If a charge
system has a size $R$, the electric potential of such system at large $r$
can be worked out as an expansion in $R/r$, with the first term
coming from the total electric charge $Q$, followed by the dipole and
quadruple potentials, etc.

Many practical quantum mechanics problems cannot be solved
without effective theory concepts. Usually, the Hilbert spaces have an
infinite dimension, so the eigenvalue problems
are almost impossible to solve exactly except for, e.g., the well-known
harmonic oscillator or hydrogen
atom. One can usually write a Hamiltonian as
\begin{equation}
  H = H_0 + H' \ ,
\end{equation}
where $H_0$ is diagonal, and $H'$ is not. Without loss
of generality, let us assume all matrix elements of $H'$
have the same size, $||H'||$. If $H_0$ contains a cluster of states that have
similar eigenvalues and span a subspace $P$ of dimension $d_P$,
then the eigenstates of $H$ with largest overlaps
with $P$ can be obtained through an effective Hamiltonian,
\begin{equation}
\label{Eq:hpert}
   H_{\rm eff} = PHP + PH'\frac{Q}{E-H_0}H'P+ ...
\end{equation}
which has a finite dimension $d_P$ and can be often diagonalized
by a computer.
$H_{\rm eff}$ has a Taylor expansion in terms of the energy ratio
$\epsilon=||H'||/\Delta E$ ($\Delta E$ is the typical energy difference between the states in $P$ and
the rest) which, when $P$ is properly chosen, could be a small parameter (if not, you are dealing with
a strongly coupled system and your luck runs out!). $P$ is sometime called model space, and the complementary
infinite-dimensional space $Q=1-P$ has been summed or ``intergated
out'' in Eq.(\ref{Eq:hpert}), as explicitly seen in the second term.
An excellent example of the above is degenerate perturbation theory, where
$P$ includes all degenerate states. When $d_P=1$, one simply
recovers the non-degenerate perturbation theory, which is ubiquitous
in quantum problems.

One important point is that
the effective wave functions obtained by diagonalizing
$H_{\rm eff}$ cannot be directly used to calculate
matrix elements of a physical observable $O$. One has to ``integrate
out'' the $Q$ contribution in $O$ to get an effective operator $O_{\rm eff}$
which will be used together with the effective wave functions. This
yields concepts such as ``effective'' charge and mass.
Thus an effective theory in quantum mechanics requires an effective Hamiltonian
and matched effective observables. The physics in the model $P$-space is presumably
simpler to understand. For example, all kinds of quasi-particles in condensed
matter physics, including anyons and Majorana zero-modes, are
effective objects arising from ordinary Coulomb interactions.
The effective theory strategy has also been widely used in solving nuclear quantum
many-body problems~\cite{Drischler:2019xuo}.

\section{What is an Effective field theory?}

By name, an EFT is a field theory with some effective degrees
of freedom (dof's), while others of the full QFT have been
integrated out. Some well-known examples are:
\begin{itemize}
\item{The standard model EFT integrates out all unknown physics
above the electroweak scale, which might be a grand unification theory
or string theory. $\epsilon=M_{\rm ew}/\Lambda_{\rm NP}$,
with $M_{\rm ew}$ as the electroweak scale and $\Lambda_{\rm NP}$ as the new physics scale.}
\item{
QCD perturbation theory (pQCD) keeps
all high-momentum dof's active, parametrizing the
physics of infrared dof's with non-perturbative
matrix elements (e.g. parton distributions). $\epsilon= \Lambda_{\rm QCD}/Q$,
where $\Lambda_{\rm QCD}$ is non-perturbative QCD scale and $Q$ is the hard scattering scale.}
\item{Chiral perturbative theory
keeps Goldstone bosons as low-energy dof's, and parametrizes the
high-energy physics in terms of ``low-energy'' (which should really be called high-energy!)
constants. $\epsilon=p/M$ where $p$ is a low momentum scale, $M$ is a hadron mass scale. }
\item{Lattice QCD is a EFT with
high-momentum dof's, $k>\pi/a$ (here $a$ is lattice spacing), integrated
out, and $\epsilon= a\Lambda_{\rm QCD}$, where $a$ is lattice spacing.}
\item{Heavy quark effective theory (HQET) considers an expansion around the quark mass $m_Q=\infty$, and
$\epsilon=\Lambda_{\rm QCD}/m_Q$. }
\end{itemize}
Thus by nature, EFT is not
so much different from the simple Taylor expansion learned
as a junior undergraduate, except there is a catch:
ultraviolet (UV) divergences!

A QFT contains infinite many dof's with low and
high momentum particles or fields needed to maintain Lorentz symmetry.
With local interactions, high-momentum dof's will produce infinite contributions to
physical observables. Since very high-momentum physics
cannot be described by the theory, they shall be integrated out and
their effects have to be parameterized in terms of some ``high-energy'' constants.
If the number of such constants is finite, the theory
is called renormlizable. Thus the renormalization
process of a QFT consists of constructing an EFT
with just low-energy dof's!
The same is in a sense true for a cut-off theory.
Therefore some would say that all QFTs are EFTs!

UV divergences, however, sometimes make Taylor expansions
not so straightforward. Let's consider a function
$f(x,\epsilon,...,\Lambda)$, which now contains a UV
cut-off scale $\Lambda$. If one does
a Taylor expansion around $\epsilon=0$, one finds
there is an ambiguity. Either you expand after finishing
the full calculation, or you take $\epsilon=0$ beforehand.
There is a difference because taking $\epsilon \to 0$ does not commute
with $\Lambda\to \infty$ and the function  $f(x,\epsilon,...,\Lambda)$
is nonanalytic at the point $\epsilon=0$! HQET in QCD is such an example.
Feynman integrals with $m_Q$ finite and with
$m_Q=\infty$ have completely different
UV behaviors~\cite{Manohar:2000dt}, which makes the quark
mass dependence of a physical quantity non-analytic!

Thus, an EFT often deals with a Taylor expansion around
a singular point of the relevant parameters, which is a bit
tricky. It requires the skill of a good graduate
student, or some smart undergraduates.

The standard EFT methodology is to take $\epsilon=0$
before doing any computation. An effective Lagrangian
is constructed to evaluate $f(x,\epsilon=0,...,\Lambda)$, and this calculation
is presumably simpler. However
this does not give the right answer $f(x,\epsilon\to 0,...,\Lambda)$.
One needs to figure out what is their difference, and
this is very important! This difference is quite
often independent of other parameters $x$. So if one does
a calculation for some specific values of $x$
and figures out the difference, the result
can be used for all $x$. Once an effective
theory calculation is done, one can get
the right Taylor series by adding up the
difference. This is called EFT matching! Matching
is needed to get the effective Lagrangian as well
as effective operators.

The UV behavior of an EFT at $\epsilon=0$ is very
different from the full theory, and this difference can be
exploited for useful purposes. It can help to sum up the
so-called large logarithms in the coupling constant expansion
of the full theory through the
renormalization-group running in the EFT. So many people working
in EFTs are doing matching and running, matching and running,
much like the life of an adult male among Yi people in China.
Again, all this is nothing but a sophisticated Taylor expansion. Armed with these
clarifications, I can talk about partons.

\section{Light-Front Effective Theory of Partons and Why It is Hard to Solve}

Partons, introduced by Feynman, are a fundamental concept in high-energy physics,
and now are a standard topic of textbooks in QFT
and high-energy physics~\cite{Ellis:1991qj,Collins:2011zzd}.
Partons are dof's in a hadron
moving at infinite momentum $P^z=\infty$, assumed along the $z$-direction.
In reality, no hadron can travel at the speed of light, even though
the proton at LHC travels at $v=0.999999999c$. Therefore,
partons are an idealized theoretical concept according to the middle
school teacher. However, without such a beautiful concept, it is hard
to imagine how to describe high-energy collisions of two protons
at LHC with thousands of particles produced!

All partons have infinite momentum (zero wavelength!), carrying a fraction of
hadron momentum, $x=\lim_{P^z\to\infty} k^z/P^z$.
They are a part of the dof's in QCD (forming a $P$-space). It is possible to single out these dof's
to write down an effective theory. The other dof's have finite momentum along
$z$ direction, which carry the zero fraction of the hadron momentum,
and hence are called ``zero modes'' ($Q$-space), including those making
up you and me!

No one had much experience working with a quantum mechanical
system travelling at the speed of light. Weinberg considered a
scalar theory in 1966 and found a set of simple rules to
do perturbation theory by eliminating the ubiquitous kinematic
infinities~\cite{Weinberg:1966jm}. After the advent of Feynman's parton model,
further studies found that Weinberg's rules can simply be reproduced by the
so-called light-front quantization~(LFQ)~\cite{Chang:1968bh,Kogut:1969xa,Drell:1970yt},
a form of dynamics proposed by Dirac as early as 1949~\cite{Dirac:1949cp}.
LFQ naturally uses the Hamiltonian formalism, bringing the EFT of partons
in a form similar to a non-relativistic many-body problem.

Many probably do not recognize that LFQ of a theory is in fact an EFT.
The effects from all finite momentum modes
cannot be directly calculated in LFQ with a
small-$x$ cut-off. There is now an infinite number
of ``low-energy'' constants that cannot be determined from
the theory itself, including all the properties of
the physical vacuum, as well as
the mass and the spin of hadrons the theory targets to
describe~\cite{Ji:2020baz}.

Solving LF quantized QCD has been exceedingly hard, if not
impossible. Ken Wilson thought of a weak-coupling approach like that
in atomic physics~\cite{Wilson:1994fk}. However, the idea has not
paid off so far. Despite much progress~\cite{Brodsky:1997de},
a systematic approximation in LFQ for the
parton structure of QCD bound states has yet to be found. Recent
progress in quantum computation has generated new hope that the problem
may ultimately be solved using quantum computers.

Inspired by the LF formalism, infrared parton modes in pQCD
are not represented by infinite momenta, but
in terms of LF correlations. More specifically,
the quark parton distribution functions (PDFs), which describe the probability
distributions of quarks, are defined through
correlation functions,
\begin{equation}
     C(\lambda)= \langle P|\overline \psi(\lambda n)\Gamma \psi(0)|P\rangle \ ,
\label{Eq:lfme}
\end{equation}
where $\psi$ is a full-QCD quark field, $n^\mu$ is a LF four-vector
$n^2=0$, $\lambda$ is the LF distance, and $|P\rangle$ is a hadron state
of any momentum. LF correlation operators (or correlators) automatically select
the parton dof's,  which in turn project the hadron state
into the effective space through the matrix elements as in Eq. (\ref{Eq:lfme}).
Therefore, partons can be defined and studied without the EFT machinery~\cite{Collins:2011zzd}.
An explicit separation of parton modes in the pQCD
Lagrangian has been made in soft-collinear effective theory
(SCET), where parton dof's are represented by LF collinear fields
~\cite{Bauer:2000yr,Bauer:2001ct,Bauer:2001yt}. Since the external hadron
states are still constructed in full QCD theory, the formulation
of SCET is different from the LFQ program where
states and operators are all manifestly in the effective space.

Imbedding the parton modes in full QCD
in terms of the collinear LF fields makes the parton physics
manifestly covariant.  Eq. (\ref{Eq:lfme}) is then amenable to
the Feynman path-integral formalism of QFT. However, it
is still impossible to calculate on a classical computer
because it involves explicitly the physical time. As such,
we say the problem is Minkowskian (the same is true
for LFQ), and it has the so-called  ``sign problem'' which is
known to be ``NP-hard'' in the language of computation theory.

I am going to argue that directly solving the
parton structure either in LFQ or in LF correlators
full QCD is actually not a good idea. The $P^z=\infty$ limit of a proton
is very similar to critical points in
condensed matter systems at which the correlation length
diverges, and infinite long-range correlations make many degrees
of freedom strongly coupled. [Correlation length $\xi$
is usually defined by an exponential behavior $\exp(-\lambda/\xi)$.]
It is hard to make theoretical
approximations as one learned from many-body systems at
critical points.

Where is the infinite-range correlation in parton physics?
The answer is at small $x$. As $x\to 0$, parton distributions
grow like $x^{-\alpha}$, where $\alpha$ is less than or equal to 1
from unitarity constraint, but generally positive.
If one Fourier-transforms this to coordinate space,
one finds the following correlation behavior,
\begin{equation}
   f(\lambda) \sim \lambda^{\alpha-1} \ ,
\end{equation}
where $\lambda$ is a conjugate variable to $x$ and the same
as the LF distance mentioned above.
Thus the correlation functions decay only algebraically with LF distance, an indication that
the system is at a critical point. In
condensed matter physics, no theorist actually tries
to understand the critical phenomena by directly
calculating at $T=T_c$!

The only known systematic approach to
solve non-perturbative QCD is
through Wilson's lattice gauge theory.
However, it only works for Euclidean QCD where
no physical time is involved. Many ideas have been
proposed to calculate parton physics on an Euclidean
lattice over the years, the most fruitful one so far is
to calculate the first few moments of PDFs which are given
by matrix elements of time-independent local operators.
However, this is far from solving the complete
parton structure.

\section{A ``New'' Parton Theory With Infinite-Momentum Hadrons}

Facing this impasse, it is useful to go back to Feynman's
original idea about partons. Feynman proposed the concept not
in the context of a QFT, but based on intuitions and experiences
with atomic physics, supplement with basic notation of relativity.
He argued that as a hadron moves at high energy or momentum, the interactions
between constituents slow down due to Lorentz
time dilation. This slowing down is dramatized by
the infinite momentum limit $P^z=\infty$, where
the hadron is now made of incoherent, non-interacting
constituents. A key information that
characterizes the state of non-interacting partons
is their longitudinal momentum distribution, $f(x)$.

Feynman arrived at PDFs from the ordinary one-dimensional
momentum distribution (other dimensions being integrated out),
$f(k^z, P^z)$ of the constituents in the hadron moving with
center-of-mass momentum $P^z$. Assuming $P^z=\infty$ is analytic,
one can Taylor-expand the momentum distribution there,
and the famous parton distribution is just the
first term of the expansion,
\begin{equation}
  f(k^z, P^z) = f(x) + f_2(x)(M/P^z)^2 + ...
\label{Eq:expansion}
\end{equation}
where $x=k^z/P^z$ and $M$ is
a hadron mass scale. The fact that the momentum distribution
of a system depends on the center-of-mass frame
is not familiar in non-relativistic systems. However,
it becomes very important when relativity is at play:
as a composite system travels faster, the internal dynamics
will change as the Hamiltonian is not invariant under
Lorentz boost.

\begin{figure}[htb]	
	\centering
	\includegraphics[width=0.6\linewidth]{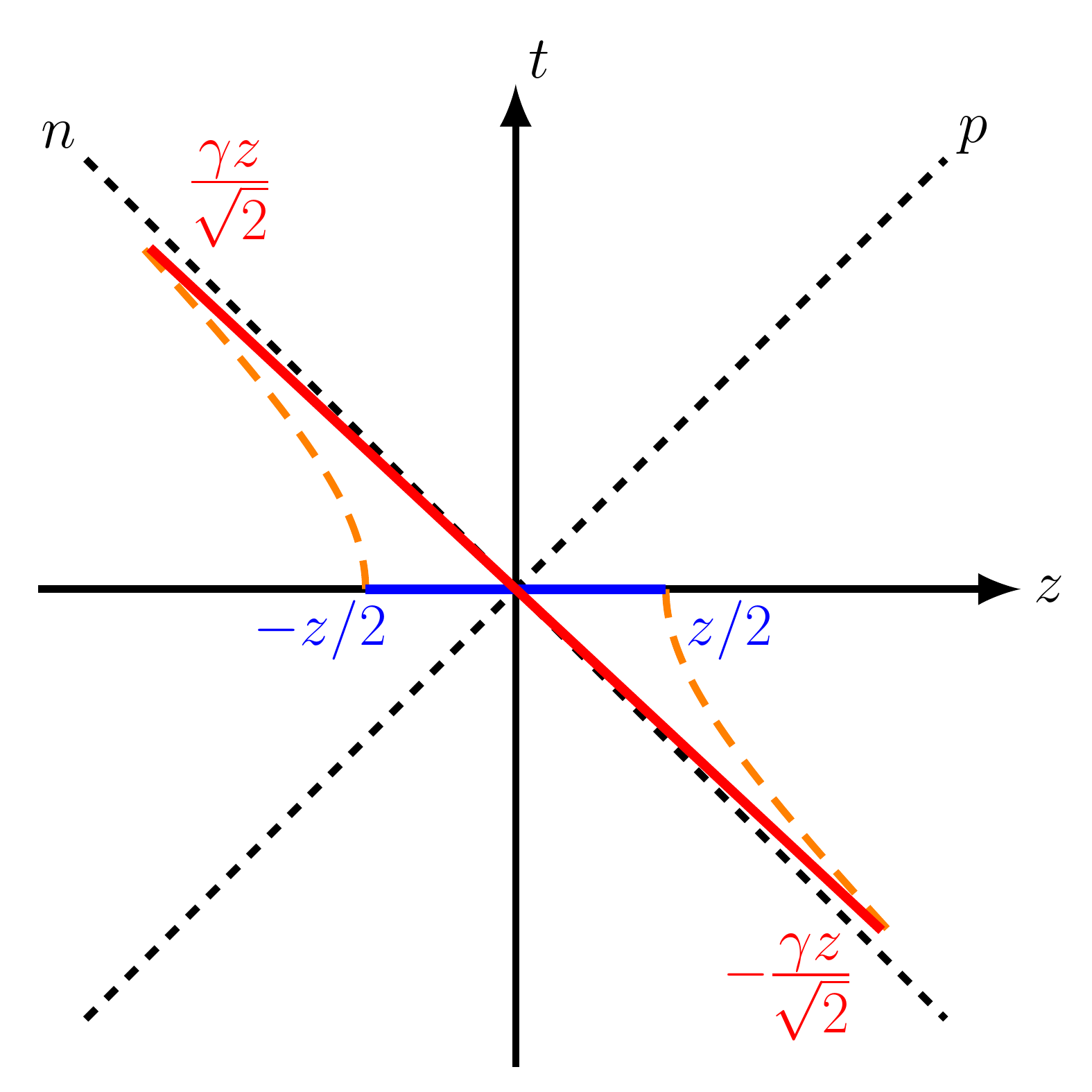}
	\caption{The connection between two pictures of parton physics: LF formalism
and Feynman infinite-momentum picture. Through Lorentz boost, the
correlation along the $z$-direction in the frame of a large-momentum hadron
is equivalent to a correlation of length $\sim \gamma z$ close to the
LF in the hadron state of zero momentum. As $\gamma\to\infty$, the latter becomes
exactly the LF correlation in Eq.(\ref{Eq:lfme}).}
	\label{fig:Boost}
\end{figure}

Feynman's original idea can be implemented in field theory
to get a ``new'' formulation of partons. For example, the quark PDFs
can now be regarded as the momentum distribution of a system
travelling with infinite momentum, and can be expressed as
Fourier transformation of the spatial correlation~\cite{Ji:2014gla},
\begin{equation}
     C(\lambda)= \langle P^z=\infty |\overline{\psi}(z)\Gamma \psi(0)|P^z=\infty\rangle \ ,
\label{Eq:Feyn}
\end{equation}
where $\lambda = \lim_{P^z\to\infty, z\to 0} (zP^z)$.
The above correlation does not involve the physical time
and can be formulated as a calculation in Euclidean field theory. It is not difficult
to see that the parton correlations in LF formalism Eq. (\ref{Eq:lfme}) and the above
differ by precisely an infinite Lorentz transformation!
The LF distance $\lambda$ in Eq. (\ref{Eq:lfme}) has a correspondence
as the infinite momentum limit of the Euclidean distance $zP^z$,
as shown in Fig. 1.

The relations between the ``old'' and ``new'' parton formulations can
further be clarified. In the LF formalims, the parton dof's are selected through
the LF collinear fields in Eq. (\ref{Eq:lfme}) and hence are intrinsically
Minkowskian. This is analogous to Heisenberg picture in
quantum mechanics, where time-dependence is carried in operators.
On the other hand, in the ``new'' representation in
Eq. (\ref{Eq:Feyn}), partons are filtered
through the infinite-momentum {\it external states} and
therefore allow an Euclidean correlation functions. This
is like Schrodinger picture in which operators are
time-independent~\cite{Ji:2020ect}.

%When I first learn the parton
%model and used $f(x)$ to calculate deep-inelastic scattering
%cross section, it worked like a charm in producing the
%scaling laws effortlessly.  However,
%I had always worried that if partons are truly
%non-interacting, the proton might fall apart. Now I know
%that in QCD, partons can slit and recombined, and can
%interact through soft quarks and gluons which cannot be
%ignored even at $P^z=\infty$, so the proton
%does not fall apart. However, these interactions do not change
%the cross sections at LHC.

Feynman, however, did not realize that $P^z=\infty$
is not well-defined in field theories with UV divergences,
and the difficulty can only be solved in asymptotically-free theory.
LF parton formalism is a result of taking $P^z\to \infty $
before one does any calculation! The resulting
parton PDFs $f(x,\mu)$ have support $-1<x<1$ (negative-$x$
corresponding to anti-particle), where $\mu$
is a renormalization scale. This is different from
a physical momentum distribution $f(y=k^z/P^z,P^z)$, which
has support $-\infty<y<\infty$ even in the $P^z=\infty$ limit.
The good news is that the difference between the limits is related to
high-momentum modes only, and they can be matched through pQCD.

The mismatch between the two $P^z\to\infty$ limits is indeed due
to UV divergences. Imagine a parton at $x=0.9$ in a proton with finite physical
momentum $P^z$, it can radiate
a gluon going backward with a momentum $x=-0.2$, and the parton ends up
with momentum $1.1P^z$. This is perfectly fine in the full theory
because no matter how large the proton momentum is, QCD always
allows a parton to carry momentum bigger than $P^z$ because the UV
cut-off is supposed to be $\gg P^z$. Where is the effect of this type
of partons in physical cross sections? It has been
taken care of through pQCD radiative
corrections. The parton EFT is supposed to include
only the low-energy scale physics, and therefore $x$ is limited
to 1 by the EFT construction.

\section{Large-momentum effective theory}

While computing $f(x)$ in the ``new'' formulation of
partons is now a Euclidean problem, it seems
still a mission impossible: we don't know how to build a
hadron state with infinite momentum. However,
Eq. (\ref{Eq:expansion}) puts the problem
again into the context of a Taylor expansion!
The idea is that one can approximate $P^z=\infty$ by a finite but
large $P^z$, and systematically correct for
any mistakes~\cite{Ji:2013dva}, keeping in mind though $P^z=\infty$
is actually singular.

The quark and gluon momentum distributions
$f(k^z,P^z)$ are quantities that can routinely be simulated
in lattice QCD for a moderately-large momentum $P^z$.
The only question is how large is a large $P^z$
that can approximate $\infty$. The answer depends
on the expansion parameter $\epsilon=(M/P^z)^2$.
One would naively expect that since $M$ is on the order of 1 GeV,
$P^z\sim 2$ GeV already gives $\epsilon=0.25$ which is already
a reasonably small parameter, a situation similar
to the charm quark when using HQET. With incoming exascale
computing, one can simulate in lattice
QCD a proton at $P^z\sim (3-5)$~GeV, making $\epsilon$
as small as $0.03$.

To make practical and systematic use of the above observation
is the main subject of {\it large-momentum effective theory}~\cite{Ji:2014gla}.
A number of important observations can be made
about this theory:

\begin{itemize}
\item{For a large momentum hadron, one can calculate
many of its static properties or correlation functions on an Euclidean lattice~\cite{Ji:2013dva}.
Besides the momentum distributions, one can also calculate transverse-momentum-dependent (TMD) distributions,
generalized momentum distributions with momentum transfer,
and static correlations of quark and gluons fields between the hadron and QCD vacuum,
etc. All of these physical properties (or quasi parton distributions)
can be used to extract the partonic physics of bound states, yielding
the generalized parton distributions (GPDs), TMDPDFs, LF wave functions, etc. ~\cite{Ji:2020ect}.}

\item{Although the naive power counting suggests the expansion
parameter to be $\epsilon=M^2/(P^z)^2$, a more careful examination
yields $M^2/(k^z)^2$, where $k^z$ is the parton momentum~\cite{Ji:2014gla,Ma:2014jla,Izubuchi:2018srq}.
Therefore, the expansion does not converge uniformly for all $x$.
Just as for an experiment for which the center-of-mass energy limits
the smallest accessible $x$, the momentum of a hadron on the lattice
limits the smallest $x$ partons one can calculate, $x_{\rm min}\sim \Lambda_{\rm QCD}/P^z$~\cite{Ji:2013dva}.
The range of a reliable LaMET calculation is $[x_{\rm min}, x_{\rm max}\sim 1-x_{\rm min}]$ which
goes to $[0,1]$ in the $P_z\to\infty$ limit.}

\item{Matching between the finite-momentum properties and parton physics
is a pQCD problem, free of infrared (IR) physics to all orders in pQCD (two
loop calculations have appeared recently~\cite{Chen:2020ody,Li:2020xml}).
The physical origin for this is that boost does not change the IR properties of a
matrix element~\cite{Ji:2014gla}. It has also been verified explicitly through pQCD analysis~\cite{Ma:2014jla}.}

\item{The momentum distribution $f(k^z,P^z)$ contains
large logarithms of $P^z$, and one can write down
an evolution equation in $P^z$~\cite{Ji:2014gla}. This momentum evolution corresponds exactly to
the renormalization group (RG) equation of PDFs apart from a simple
field renormalization. Thus the parton RG evolution has its physical
origin as the change of momentum distributions due to the change
of Hamiltonian and states under Lorentz boost, and helps to
sum large-$P^z$ logarithms in $f(k^z,P^z)$.}

\item{Higher order (power) corrections in the Taylor expansion can be worked out systematically,
and can be calculated through lattice simulations as well. They help to improve
the precision of a LaMET calculation.}

\item{In critical phenomena, one can have nominally different physical
systems sharing the same critical-point properties. Examples include
the spontaneous magnetization and liquid-gas
critical point. Similarly, in LaMET, one can use
different Euclidean operators to get the same PDFs~\cite{Hatta:2013gta}.
This is guaranteed by the projection through large-momentum
external states and is a universality phenomenon.
Therefore, apart from the static physical distributions, one can also calculate
PDFs using many other Euclidean correlations including
current-current correlators, etc.~\cite{Braun:2007wv,Ma:2017pxb}.}

\item{LaMET provides a general recipe to calculate
light-like correlations entering the factorizations
of high-energy processes. One can replace the
light-like Wilson lines in operators by large-momentum hadron
external states when appropriate, and
if the result is time-independent, it
can be calculated by a lattice simulation. An excellent example is the soft
function appearing in TMD factorization~\cite{Ji:2019sxk}.}
\end{itemize}

\begin{figure}[htb]	
	\includegraphics[width=0.7\linewidth]{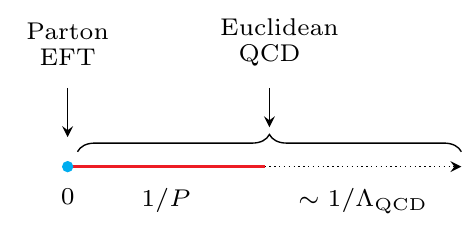}
	\caption{The parton EFT is an infinite-momentum limit of Euclidean QCD after proper matching.
Solid line shows the region of convergence of the LaMET expansion. Similar to critical phenomena,
the parton limit $P^z=\infty$ is like a critical point and corresponds to a fixed point of the momentum
renormalization group. The solid line represents the critical region or the ``basin of attraction'' of the
renormalization flow.}
	\label{fig:PDFS}
\end{figure}

Having discussed how the LaMET formalism actually works, I am in the
position to discuss the nature of this approach in light of the properties of
EFTs discussed in Sec. III.

\begin{itemize}
\item{Approximating an infinite momentum by a finite momentum in LaMET
is not new, and is exactly what has been used in lattice QCD. The highest momentum
on a lattice is $\pi/a$, which is supposed to be taken to infinity in the end.
One can handle this limit through a systematic expansion in
$a\Lambda_{\rm QCD}$.}
\item{Going back to Eq.(1), one can invert the expansion,
through
\begin{eqnarray}
    f(x,0,0,...) &&= f(x,\epsilon,\delta,...)- \epsilon f_\epsilon(x,\epsilon,\delta,...)\nonumber\\
        && -\delta f_\delta (x,\epsilon,\delta,...)+... \ .
\end{eqnarray}
Now one can regard $f(x,\epsilon,\delta,...)$ as an effective description
of $f(x,0,0,...)$! Therefore, the magic word ``effective'' is not absolute, but
mutual, analogous to a mirror symmetry. In one sense, LaMET
uses calculations in the full QCD Lagrangian, right-hand side in Eq. (8),
to {\it simulate} the parton physics on the left, and the
QCD on the lattice is an effective theory for the LF theory of parton.}
\item{On the other hand, one may regard the emergent parton physics
as an EFT describing the physical properties of hadrons at large $P^z$ in full QCD~\cite{Ji:2014gla}.
This interpretation is analogous to HQET which uses an infinite-heavy quark
to describe the physics of a heavy quark. Whereas the usual effective theories
use matching and running to get the effective Lagrangian and effective operators and then use
them to calculate, LaMET does
matching and running for {\it all} physical observables.}
\item{Following Sec. V, $P^z=\infty$ is like a critical point
in condensed matter systems. At critical points,
long wavelength modes dominate, and the critical phenomena
are studied through systems very close to, but
not exactly at, $T_c$. LaMET follows the same spirit~\cite{Ji:2014gla}, as shown in Fig. 2.
The correlation length $\xi$ is finite in the critical region, and is proportional to
$P^z/\Lambda_{\rm QCD}$.}
\end{itemize}

For all these reasons, the name LaMET has been given to
the strategy of solving parton physics through simulations
of Euclidean QCD at a large hadron momentum
It is an effective theory for parton structure
without directly dealing with the parton dof's themselves,
like in LF quantization.

\section{Instead of A Conclusion: To Control or Not to Control?}

Partons provide a powerful language which is used
everyday by thousands of physicists to describe
high-energy collisions. Although conceptually
simple, the mathematics of describing parton
interactions forming the high-energy proton
is very challenging. Therefore, it is
important that any approach to calculating
PDFs must have systematic controls of errors.

The infinite-momentum-state representation in Eq. (\ref{Eq:Feyn})
provides a starting point to apply
approximation methods. LaMET formalism
begins with lattice QCD data generated with large hadron
momenta, and offers a method to extract parton
structure through a sophisticated Taylor expansion.
Since a most important hallmark of an EFT
is power counting, i.e., uncertainties can be quantified
in sizes of expansion parameters, LaMET meets
this criteria as an EFT, although calculating the power
correction is by no means a small task.

\begin{figure}[htb]	
	\includegraphics[width=0.99\linewidth]{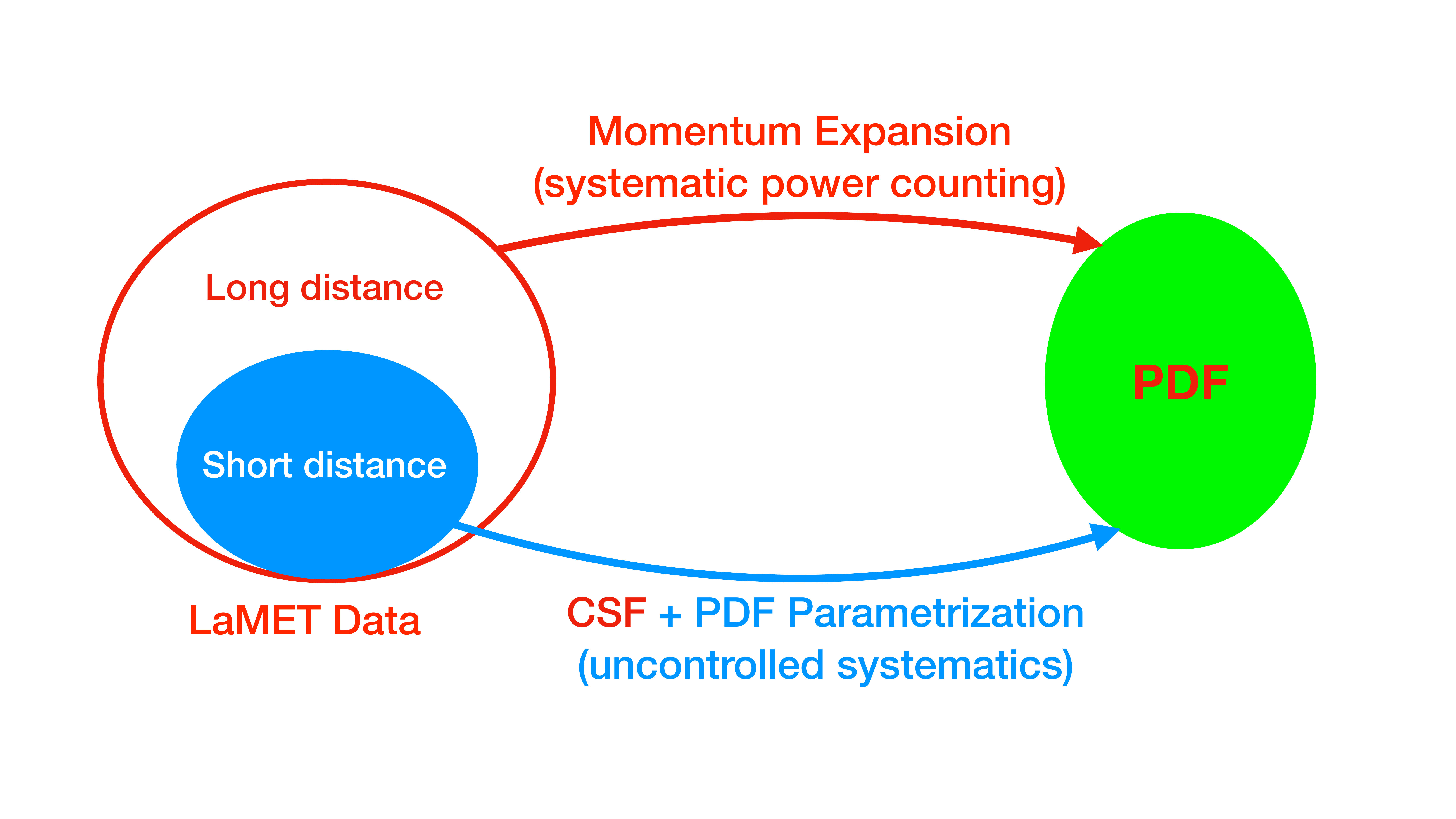}
	\caption{Comparison between the momentum and coordinate
expansions to analyze LaMET data. The latter
can be applied only to a subset of parton observables, for which
the concept of a short-distance expansion can be applied.}
	\label{Fig:compare}
\end{figure}

The Euclidean correlator in Eq. (\ref{Eq:Feyn}) introduced in Ref. \cite{Ji:2013dva}
has also been considered in coordinate-space factorizaton (CSF)~\cite{Radyushkin:2017cyf},
introduced in an early work on  meson distribution amplitude with current-current correlator
\cite{Braun:2007wv}, see also~\cite{Ma:2017pxb}.
The correlator can be factorized in terms of the LF correlations
with an expansion parameter $(z\Lambda_{\rm QCD})^2$. The formalism
naturally suits for calculating the moments of PDFs or short-distance
LF correlations. To obtain the full parton physics, however, one has to
simultaneously consider the constraint on the external momentum,
\begin{equation}
          \lambda = zP^z\sim 1, ~~~~~~~~ P^z\sim 1/z \gg \Lambda_{\rm QCD}\ .
\end{equation}
This is identical to the observation in Ref. \cite{Ji:2014gla}: one must use
large momenta to capture the full dynamical range of PDFs, which requires information on
long-range correlations in $\lambda$. Despite complete equivalence~\cite{Ji:2017rah,Izubuchi:2018srq},
there is a tendency in the literature to translate every momentum-expansion paper under
the Sun into an equivalent CSF form, although some analytical matching calculations might
be conveniently done in coordinate space. Not surprisingly, the same LaMET
lattice data are needed for a CSF analysis to get PDFs.
[Nominally, CSF can also admit data at small $P^z$, but the same information is
already in large $P^z$ data at smaller $z$.]

Interestingly, however, when the CSF methodology is applied to
large-$P^z$ data, it has not yet produced a controlled expansion scheme for
$x$-dependent PDFs. The CSF re-interpretation of the correlation
functions prohibits using large-$z$ data, and the Fourier transformation to
momentum space becomes incomplete due to a limited $P^z$.
As a consequence, one has to parametrize the physical PDFs and to covert
them to coordinate space in order to make fits to lattice data at
$z\ll 1/\Lambda_{\rm QCD}$. The process generates uncontrolled systematics
through model parametrization and fitting. Alternatively, one
might try to model the high-order $(z\Lambda_{\rm QCD})^2$ contributions
and subtract them at large-$z$. Again such modeling will introduce
uncontrolled systematics. In reality, short distance expansion for
large-$P^z$ matrix elements is unnecessary, the momentum expansion
automatically quantifies the high-order power contributions
in the large-$z$ data through Fourier analysis. The relationship between
momentum and coordinate-space analysis of the
LaMET data is shown in Fig.~\ref{Fig:compare}.

This seems a bit strange at first, because ordinarily one
would expect so long at $P^z$ is large enough, bulk of the parton
physics shall reproduced by short distance data. From the above
discussions, however, it is clear that physics and power counting
systematics naturally lead to a momentum expansion,
not CSF which unavoidably introduces unnecessary cuts
on LaMET data. In more sophisticated
LaMET applications such as TMDPDFs and LF wave functions, the
coordinate space correlation data may not have an apparent CSF interpretation,
and the momentum expansion might be the only
game left to play~\cite{Ji:2020ect}.

The major difference between models and EFT's is about control of systematics.
In 1998, H. Georgi was invited to a conference at JLab,
``Quark Confinement and the Hadron Spectrum III''.
He gave an after dinner talk in which he listed 10 reason for
attending the meeting. Reason No. 7 says, ``you are
not a control freak, so you don't like controlled approximations.''
He went on saying, ``The difference between the nuclear physics
and particle physics traditions always amazes me here. Trained
as a particle physicist, I tend to like to control my approximations
even if it means distorting the physics. A lot of the speakers
here clearly didn't care. I don't know whether this is good or bad.''
I had been educated as a nuclear theorist by an excellent Ph.D. advisor,
the title of my thesis was ``Shell-Model Effective Interactions for N=50 Nuclei.''
So I had been trained as a control freak!

\section*{Acknowledgments}

I thank J. W. Chen, Y. Z. Liu, J. P. Ma, A. Sch\"afer, W. Wang, B. W. Xiao, F. Yuan, J. H. Zhang and Y. Zhao for
helpful discussions and comments related to the subject of this paper. I particularly thank
Y. Q. Ma for many discussions and correspondences helping to make the paper
readable to non-experts, and A. Sch\"afer for a careful reading and editing suggestions of the manuscript.
I also appreciate the figures helped by Y. Zhao.
This work is partially supported by the U.S. Department of Energy under Contract No. DE-SC0020682, and
by Center for Nuclear Femtography operated by Southeastern University Research Association in Washington DC.

\bibliography{references}

\end{document}